\documentclass[12pt,]{article}
\usepackage{lmodern}
\usepackage{amssymb,amsmath}
\usepackage{ifxetex,ifluatex}
\usepackage{fixltx2e} 
\ifnum 0\ifxetex 1\fi\ifluatex 1\fi=0 
  \usepackage[T1]{fontenc}
  \usepackage[utf8]{inputenc}
\else 
  \ifxetex
    \usepackage{mathspec}
  \else
    \usepackage{fontspec}
  \fi
  \defaultfontfeatures{Ligatures=TeX,Scale=MatchLowercase}
\fi
\IfFileExists{upquote.sty}{\usepackage{upquote}}{}
\IfFileExists{microtype.sty}{%
\usepackage{microtype}
\UseMicrotypeSet[protrusion]{basicmath} 
}{}
\usepackage[margin = 1in]{geometry}
\usepackage{hyperref}
\hypersetup{unicode=true,
            pdftitle={State Drug Policy Effectiveness: Comparative Policy Analysis of Drug Overdose Mortality},
            pdfauthor={Jarrod Olson, MPP; Po-Hsu Allen Chen, PhD; Marissa White, MPH; Nicole Brennan, PhD (Battelle Memorial Institute); and Ning Gong, PhD (Legal Sciences, LLC)},
            pdfborder={0 0 0},
            breaklinks=true}
\usepackage{longtable,booktabs}
\usepackage{graphicx}
\IfFileExists{grffile.sty}{%
\usepackage{grffile}
}{}
\makeatletter
\def\maxwidth{\ifdim\Gin@nat@width>\linewidth\linewidth\else\Gin@nat@width\fi}
\def\maxheight{\ifdim\Gin@nat@height>\textheight\textheight\else\Gin@nat@height\fi}
\makeatother
\setkeys{Gin}{width=\maxwidth,height=\maxheight,keepaspectratio}
\IfFileExists{parskip.sty}{%
\usepackage{parskip}
}{
\setlength{\parindent}{0pt}
\setlength{\parskip}{6pt plus 2pt minus 1pt}
}
\ifx\paragraph\undefined\else
\let\oldparagraph\paragraph
\renewcommand{\paragraph}[1]{\oldparagraph{#1}\mbox{}}
\fi
\ifx\subparagraph\undefined\else
\let\oldsubparagraph\subparagraph
\renewcommand{\subparagraph}[1]{\oldsubparagraph{#1}\mbox{}}
\fi

\let\rmarkdownfootnote\footnote%
\def\footnote{\protect\rmarkdownfootnote}

\usepackage{titling}

  \title{State Drug Policy Effectiveness: Comparative Policy Analysis of Drug Overdose Mortality}
    \pretitle{\vspace{\droptitle}\centering\huge}
  \posttitle{\par}
    \author{Jarrod Olson, MPP\footnote{Corresponding author, \href{mailto:olsonjr@battelle.org}{\nolinkurl{olsonjr@battelle.org}}}; Po-Hsu Allen Chen, PhD; Marissa White, MPH; Nicole Brennan, PhD (Battelle Memorial Institute\footnote{505 King Ave, Columbus, OH 43201}); and Ning Gong, PhD (Legal Sciences, LLC\footnote{WeWork c/o Legal Science 1601 Market Street, Fl. 19 Philadelphia, PA 19103})}
    \preauthor{\centering\large\emph}
  \postauthor{\par}
    \date{}
    \predate{}\postdate{}
  
\usepackage{booktabs}
\usepackage{longtable}
\usepackage{array}
\usepackage{multirow}
\usepackage{wrapfig}
\usepackage{float}
\usepackage{colortbl}
\usepackage{pdflscape}
\usepackage{tabu}
\usepackage{threeparttable}
\usepackage{threeparttablex}
\usepackage[normalem]{ulem}
\usepackage{makecell}
\usepackage{xcolor}

\usepackage{enumerate} \usepackage{graphicx} \usepackage{booktabs} \usepackage{pdflscape} \usepackage{caption} \usepackage{xcolor} \usepackage{setspace} \usepackage{float} 

\begin{document}
\maketitle
\begin{abstract}
Opioid overdose rates have reached an epidemic level and state-level policy innovations have followed suit in an effort to prevent overdose deaths. State-level drug law is a set of policies that may reinforce or undermine each other, and analysts have a limited set of tools for handling the policy collinearity using statistical methods. This paper uses a machine learning method called hierarchical clustering to empirically generate ``policy bundles'' by grouping states with similar sets of policies in force at a given time together for analysis in a 50-state, 10-year interrupted time series regression with drug overdose deaths as the dependent variable. Policy clusters were generated from 138 binomial variables observed by state and year from the Prescription Drug Abuse Policy System. Clustering reduced the policies to a set of 10 bundles. The approach allows for ranking of the relative effect of different bundles and is a tool to recommend those most likely to succeed. This study shows that a set of policies balancing Medication Assisted Treatment, Naloxone Access, Good Samaritan Laws, Medication Assisted Treatment, Prescription Drug Monitoring Programs and legalization of medical marijuana leads to a reduced number of overdose deaths, but not until its second year in force.
\end{abstract}

\hypertarget{introduction}{%
\section{Introduction}\label{introduction}}

Between 1999 and 2017, the total number of opioid related deaths in the U.S. increased more than 509 percent (Centers for Disease Control and Prevention, 2018). In an effort to slow the ``epidemic level'' rise in mortality rates, law makers have developed a myriad of policies targeting prevention, treatment and judicial consequences.

Since 1999, Legal Science (2018) has cataloged state-level drug policies into the Prescription Drug Abuse Policy System (PDAPS) across 50 states. PDAPS codes specific laws in force in a state as a policy and this paper uses this same narrow definition of policy. Between 2000 and 2016 there has been a 2092 percent increase in state-level drug-related policy changes (i.e.~implementing or repealing a policy). The opioid epidemic has impacted states at varying times and to varying degrees resulting in segmented policies with variation from state to state. As of 2016, each state had an average number of 33.25 policies targeting the reduction of opioid-related deaths. While federal drug policy still reflects a heavy focus on criminalization, these state-level policies focus on a wide range of topics, from marijuana legalization, to mandatory sentencing for opioid distribution.

With this much variation in the policy system, decisionmakers are reasonably challenged to understand which policy, or combination of policies, to implement to minimize drug overdose deaths. With more than 145 policies nationwide there are \ensuremath{8.0479261\times 10^{251}} unique policy combinations to select from. The comparative policy analysis (CPA) literature has attempted to address this complex environment, but existing methods are inadequate to prioritize policies in such a rapidly changing environment. Qualitative CPA methods are limited in their ability to generalize, while quantitative CPA studies must limit their investigation to only one or a small sub-set of policies to allow for statistical testing, eliminating the ability to prioritize.

This paper identifies and explicitly estimates the relative effect of policy clusters composed of sets of policies implemented together in states over time. It utilizes an existing data set (PDAPS), a well-founded methodology drawn from quantitative CPA, and a new algorithm to assess and prioritize the policy interventions employed in the recent history of state-level drug policy intervention. We hypothesize that sets of policies, generated empirically using hierarchical clustering based on the time and place they were implemented together, will change the drug overdose death rate. We will test this hypothesis using a cross-sectional, interrupted time-series analytical approach. This approach blends work from the qualitative CPA discipline with quantitative approaches.

\hypertarget{background}{%
\section{Background}\label{background}}

The Centers for Disease Control and Prevention (2018) calls the current opioid overdose crisis an epidemic. Opioids are the leading cause of death for people under the age of 50, and about 130 Americans die every day from an opioid overdose. The CDC has identified three ``waves'' of drug overdose deaths: increased prescription of opioids in the 1990s with deaths increasing since at least 1999 at a steady rate, the growth of heroin starting in about 2010, and the sudden climb in deaths caused by synthetic opioids such as Fentanyl starting around 2013.

\begin{figure}
\centering
\includegraphics{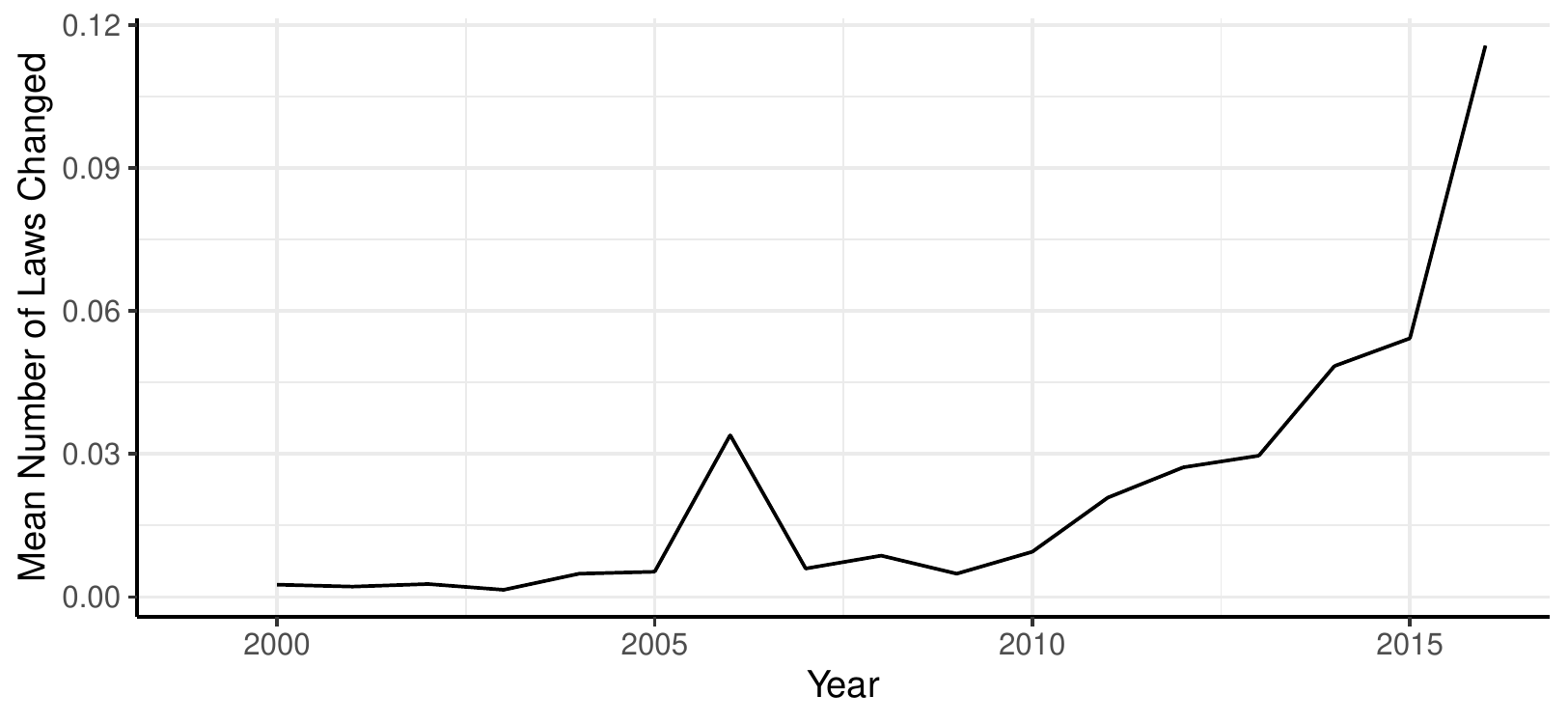}
\caption{\label{fig:pdapsGrowth}Mean Number of State Drug Laws Changed, by State.}
\end{figure}

Specific policy approaches have developed to address each wave and state-level innovations have reacted with rapid growth and massive churn in state-level policy since 2000, as shown in Figure \ref{fig:pdapsGrowth}. These innovations include the development of Prescription Drug Monitoring Programs (PDMP) to address patient-specific prescribing behavior, and policies to address the rising death rate attributed to heroin and Fentanyl through Naloxone Access Laws (NAL) and Good Samaritan Laws (GSL), which both aim to respond to actual overdoses and prevent deaths. Each policy addresses different causal issues associated with drug overdose. PDMP increases scrutiny on the supply of opiates, typically by creating a registry to track prescriptions. NAL improves access to medication that reverse active opioid overdoses, and GSL reduces the risk for persons reporting a drug overdose (who might otherwise be arrested for engaging in a criminal act like consuming opiates with the person experiencing an overdose). Haegerich, Paulozzi, Manns, \& Jones (2014) found that there is very limited reliable data about drug policy effectiveness, noting that ``the quality of evidence for the impact of state legislation on provider behavior, patient behavior, and health outcomes is low,'' and that without improvements and studies of effectiveness, states, regulatory agencies, and organizations will be unable to make informed decisions.

In response to this innovation in drug policymaking, a number of studies have been conducted to understand which policies work to reduce drug overdose deaths, and the context around it. Several authors have attempted studies that either articulate every minimal difference in laws (as passed), often leading to a narrow qualitative case study approach (see Ritter, Livingston, Chalmers, Berends, \& Reuter, 2016 for several). The literature on these qualitative case studies often does not generalize, but can help answer questions about the policy implementation and hypothesize the conditions that facilitated or blocked success (Ritter et al., 2016). Other authors have attempted to find more precise and generalizable estimates of overall effect using quantitative methods, or mixed-methods that utilize qualitatively defined ``sets'' of policies (e.g.~Blackman, Wistow, \& Byrne, 2013; Devers, Lallemand, Burton, Zuckerman, \& Authors, 2016; and Pardo, 2017). Policies that are identified and used in quantitative papers often rely on a summative index score, potentially derived from empirical methods, data-driven policy coding (such as the data set produced by Legal Science), differences based on geography (Keele \& Titiunik, 2015), or described policy differences that can be coded for analysis. These approaches typically rely on a variety of statistical methods (Ritter et al., 2016).

Each policy implemented in a state has specific details distinguishing it from other policies that are critical for comparing effectiveness. These differences are often ignored or treated minimally by controlling for the jurisdiction as the intervention, or all of the policies within a jurisdiction, leading to ``a hazy definition of policy that leads to confusion in the specification and measurement of the phenomena being studied'' (Burris, 2017). Burris (2017) propose ``the use of transparent, reliable method to produce detailed observations of the apparent characteristics of policy that can be used in evaluation research.'' An individual policy does not define the entire legal environment that might affect outcomes. As an illustration, studying NAL in isolation will miss the critical importance of GSL as an enabler. GSLs address the problem that individuals can be arrested for illegal possession and use of opioids if they call 911 for a friend experiencing an overdose, thus disincentivizing a person from calling for help, where Naloxone might be distributed broadly (i.e.~first responders).

Finally, in both qualitative and quantitative methods, there is often a lack of theory applied to analysis. Baptist, Carrie \& Befani, Barbara (2015) note that theory is a critical tool for triangulating findings, or making theory-based qualitative evaluations, and is a critical ingredient in so-called set-based methods for classifying data into groups for analysis. In quantitative methods, a strong theory of change is critical to understand the parameters of the analysis, as well as identify a spurious finding. A theory of change also guides the scientific questions asked over the course of the study (Burris et al., 2010).

\hypertarget{methods}{%
\section{Methods}\label{methods}}

CPA is a broadly defined discipline for theory-driven comparisons of policy effectiveness. It has no singular set of methodological tools but utilizes the most appropriate analytical method for the policy under assessment. In all cases, CPA aims to explain the relationship between policies and their effects (Ritter et al., 2016). Necessarily, this includes a specified outcome and a given policy, studied over two or more jurisdictions. In Table \ref{tab:cpaCharacteristics} the characteristics of this study are described using Ritter et al. (2016)'s framework.

\begin{table}

\caption{\label{tab:cpaCharacteristics}CPA Characteristics as defined by Ritter, 2016}
\centering
\begin{tabular}[t]{l|>{\raggedright\arraybackslash}p{2in}|>{\raggedright\arraybackslash}p{2in}}
\hline
Characteristic & Definition & This Study\\
\hline
Jurisdiction & Legal entity used as unit of analysis & 50 States\\
\hline
Treatment of Time & How is time used in the comparison & Lagged independent variable with policy shock\\
\hline
Purpose & Study's purpose & Identify highest impact sets of policies on reducing drug overdose deaths\\
\hline
Method & Quantitative or Qualitative & Quantitative\\
\hline
Comparative Design & How is comparison operationalized & Comparing concurrent state policy implementation\\
\hline
Policy Specification & How is the policy operationalized & Drug-related policies clustered together as "bundles"\\
\hline
\end{tabular}
\end{table}

Our specification follows from Burris et al. (2010), which frames the role of public health law on health outcomes.Laws, which are ``interventional,'' ``infrastructural,'' or ``incidental'' change legal practice, which affects both changes in behavior directly, and changes in environment that indirectly lead to additional changes in behavior. These changes in environment and behavior yield changes in population health. In this case, viewing drug overdose deaths as a public health problem suggests that all corresponding changes in legal practice will affect the drug overdose rate, including changes in drug enforcement (interventional), the addition of new infrastructure to support drug enforcement (infrastructural), or policies that affect health as a secondary feature (incidental).

\begin{figure}[H]
\includegraphics[width=450px]{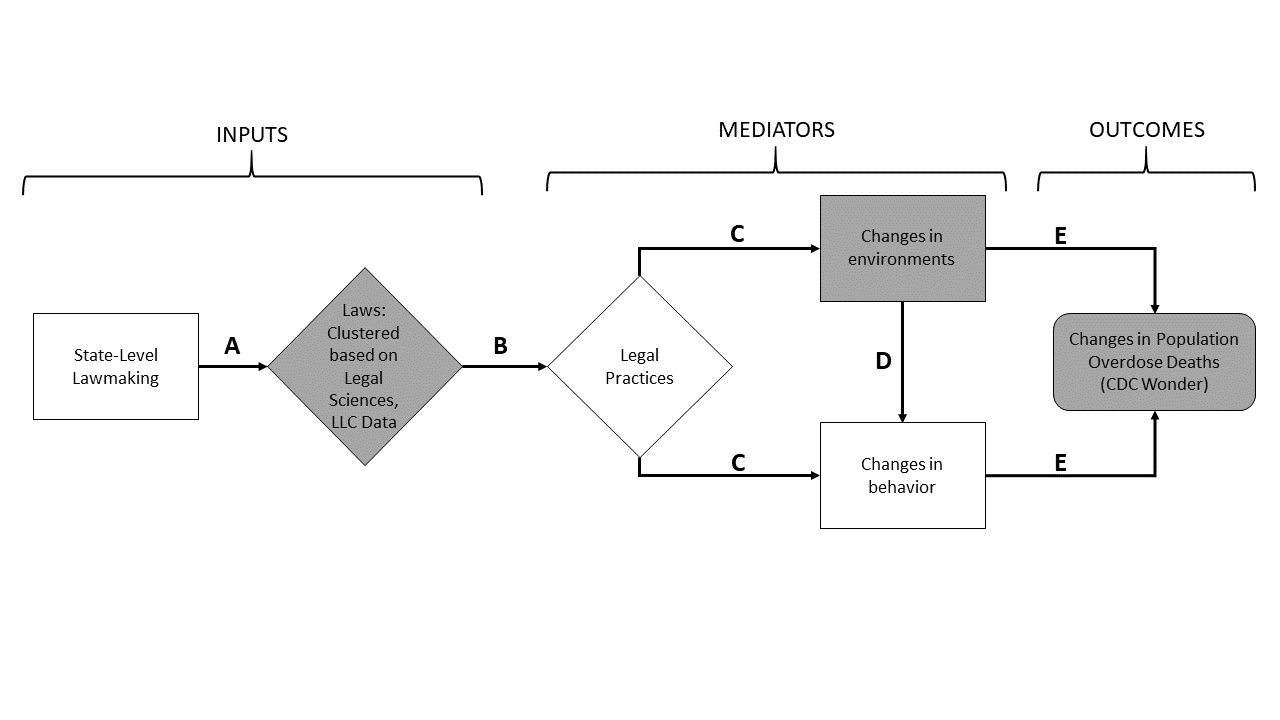} \caption{Public Health Legal Research Framework, Applied to this Analysis.}\label{fig:PLHRApplied}
\end{figure}

Figure \ref{fig:PLHRApplied} shows the theoretical model from Burris et al. (2010) modified to represent this study. We are directly modeling sets of laws, changes in environments (through a set of control variables), and the outcome (drug overdose deaths), i.e.~B-C-D-E. We expect laws that reduce drug overdose deaths through environmental changes that also affect behavioral changes to reduce drug overdoses the most. However, many environmental changes could yield a behavior change that has a mitigating effect. For example NALs aim to reduce drug overdose deaths by access to Naloxone for people experiencing an opioid related drug overdose. One potential unintended consequence may be that this environmental change could increase drug use, and overdoses if the population believes an overdose can be easily counteracted.

We specify our policies using categorical classifications of individual policies that are not mutually exclusive, and are often implemented within a system of policies that are both complementary and conflicting. Those specifications are converted into bundles based on the states that implemented them at the same time. As an example, if policy one and policy two are implemented in states x and y in the same years, but policy three is only implemented in state z in a different year, then there are two bundles, one for policies one and two, and one for policy three.

This paper uses a time-series, cross-sectional method with an interrupted time-series design as described by Bernal, Cummins, \& Gasparrini (2017) to estimate the effect of policy bundles on drug overdose deaths. The intervention (\(intervention\)) is configured as a mutually exclusive dummy variable, with panel data of states (\(s\)) over time (\(t\)). Using a fixed-effects regression (segmentation), where time-invariant factors of each observation (\(s\)) are controlled using a dummy for the unit of analysis (i.e.~state-years). Correlation and temporal order are necessary to establish causality, so the policy clusters are lagged by one year. Equation \eqref{eq:itsGeneral} shows a general interrupted time-series design using panel data and an intervention.

\begin{align}
outcome_{ts} = \beta_{0} + \beta_{t} time + \beta_{s} unitOfAnalysis + \beta_{k} control_{sk(t)} + \notag \\
  \beta_{i} intervention_{s(t-1)} + \beta_{deaths(t-1)} + \epsilon \label{eq:itsGeneral}
\end{align}

In this case, the ``intervention'' is the policy bundle in force within a given state. Drawing from Burris et al. (2010), this paper includes a set of sociodemographic and environmental factors as controls. Because drug overdoses deaths are a count variable, highly dependent on population, the value is log transformed and normalize by population in a Poisson regression as shown in Equation \eqref{eq:itsSpecific}.

\begin{align}
drugOverdoseDeathsPerCapita_{ts} = \beta_{s} State + \beta_{1} \log(Population_{ts}) + \beta{2} Gender_{ts} + \notag \\
  \beta_{3} PrescribingRate_{ts} + \beta_{4} Gini_{ts} + \beta_{5} Income_{ts} + \beta{6} Deaths_{s(t-1)} + \notag \\
  \beta_{c} policyCluster_{st-1)} + \epsilon \label{eq:itsSpecific}
\end{align}

Variables for the equation are drawn from a variety of publicly-available, state-level datasets and the \(policyCluster\) variable is constructed for this analysis from the PDAPS datasets. Although policy and overdose data are available from 1999 forward, data analysis is limited to 2006 - 2016 when most of the opioid-related policy changes started. Table \ref{tab:dataCharacteristics} shows summary information about each variable.

\begin{longtable}[]{@{}lllrrr@{}}
\caption{\label{tab:dataCharacteristics}Data characteristics for analysis variables, n observation = 1122.}\tabularnewline
\toprule
Variable & Type & Years & n & Mean & SD\tabularnewline
\midrule
\endfirsthead
\toprule
Variable & Type & Years & n & Mean & SD\tabularnewline
\midrule
\endhead
\(drugOverdoseDeaths\)\footnote{Centers for Disease Control and Prevention (2017)} & Count & 2006-2016 & 1122 & 339.14 & 408.60\tabularnewline
\(Population\)\footnote{Centers for Disease Control and Prevention (2017)} & Count & 2006-2016 & 1122 & 3051925.43 & 3425227.85\tabularnewline
\(PrescribingRate\)\footnote{Disease Control and Prevention (2019)} & Rate & 2006-2016 & 1122 & 80.96 & 22.98\tabularnewline
\(Gini\)\footnote{Bureau (2017b)} & Continuous & 2006-2016 & 1122 & 0.46 & 0.02\tabularnewline
\(Income\)\footnote{Bureau (2017a)} & Continuous & 2006-2016 & 1122 & 52.53 & 8.81\tabularnewline
\(policyCluster\)\footnote{Generated in this analysis from Legal Science (2018)} & Multinomial & 2006-2016 & 1122 & NA & NA\tabularnewline
\(Gender\)\footnote{Centers for Disease Control and Prevention (2017)} & Binomial & 2006-2016 & 1122 & NA & NA\tabularnewline
\bottomrule
\end{longtable}

\hypertarget{outcome-data-cdc-wonder-data}{%
\subsection{Outcome Data: CDC Wonder Data}\label{outcome-data-cdc-wonder-data}}

Wide-ranging Online Data for Epidemiologic Research (WONDER) (Centers for Disease Control and Prevention, 2017) is an online database that makes several CDC datasets available to public health professionals. This paper includes the drug-overdose death rates and prescribing data. WONDER suppresses the number of deaths if there are fewer than 10 deaths in a query to protect personal privacy. To reduce the amount of censored data, yearly instead of monthly age-adjusted drug-related overdose underlying cause of death counts are used for all 50 states and the District of Columbia (DC) by gender between 1999 to 2016 for a total of 1,836 observations, where 15 (0.8 percent) censored data were imputed by five (median of 0-10). Yearly state-level population data were also obtained from WONDER.

\begin{figure}
\centering
\includegraphics{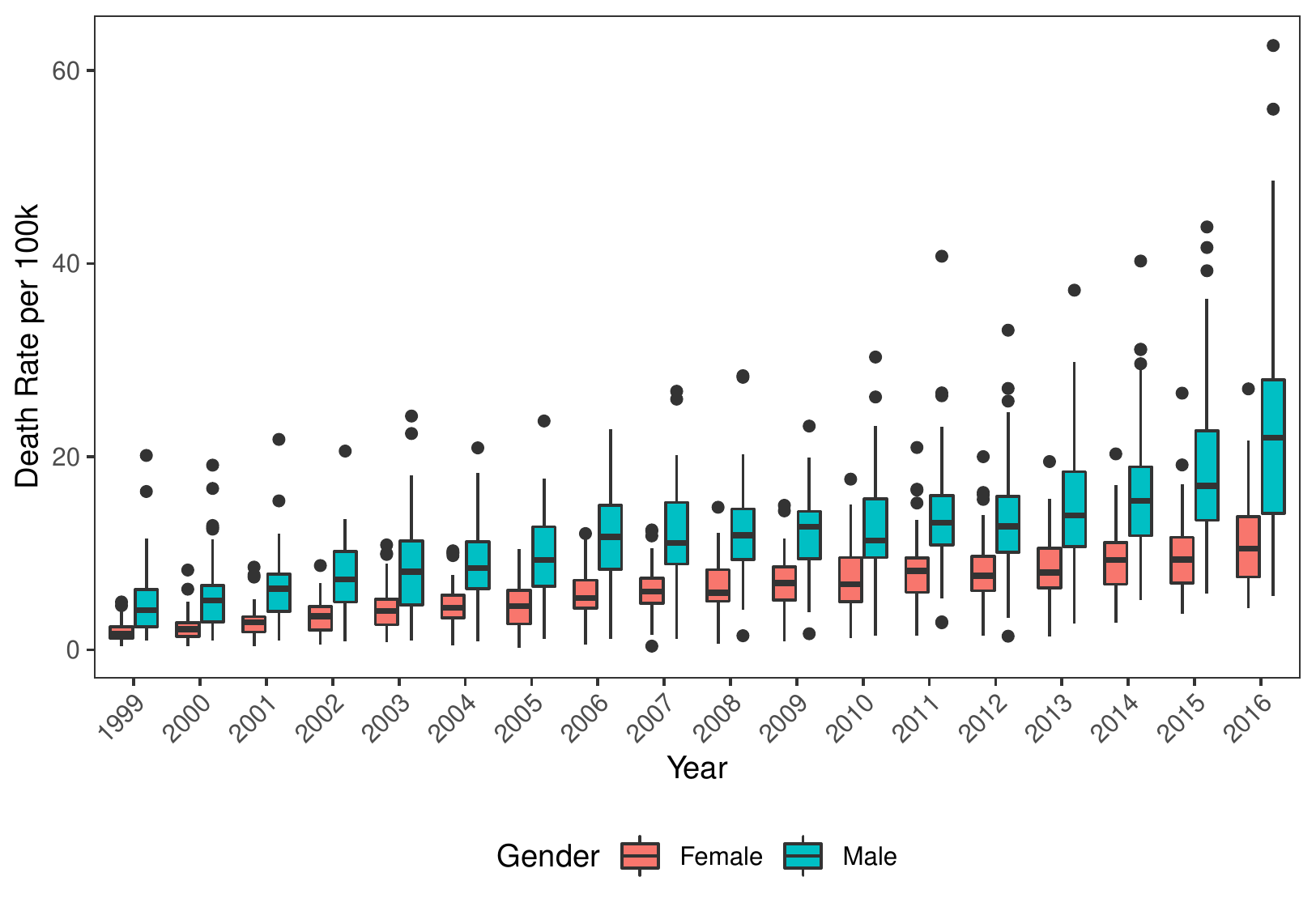}
\caption{\label{fig:WONDER}Death Rate Per 100,000 Population.}
\end{figure}

Figure \ref{fig:WONDER} shows the box plots of drug-related overdose death rates (number of deaths divided by the population size) reported in WONDER from 1999 to 2016 across 50 states and DC by gender. On average, the drug-related overdose death rates have increased 513 percent between 1999 and 2016, and the slope of the trend has become steeper in recent years. During 1999 to 2016, West Virginia (17.3), Kentucky (15.5), and Nevada (14.2) had the highest average drug overdose death rates, while Maryland (2.2), North Dakota (2.8), and South Dakota (3.0) had the lowest average death rates. Rates of drug-related overdose death are significantly higher for males than females. The male overdose rate increased from 5.1 per 100,000 population in 1999 to 24.3 while the rate for female increased from 1.9 to 11.3.

\hypertarget{pdaps}{%
\subsection{PDAPS}\label{pdaps}}

\begin{figure}
\centering
\includegraphics{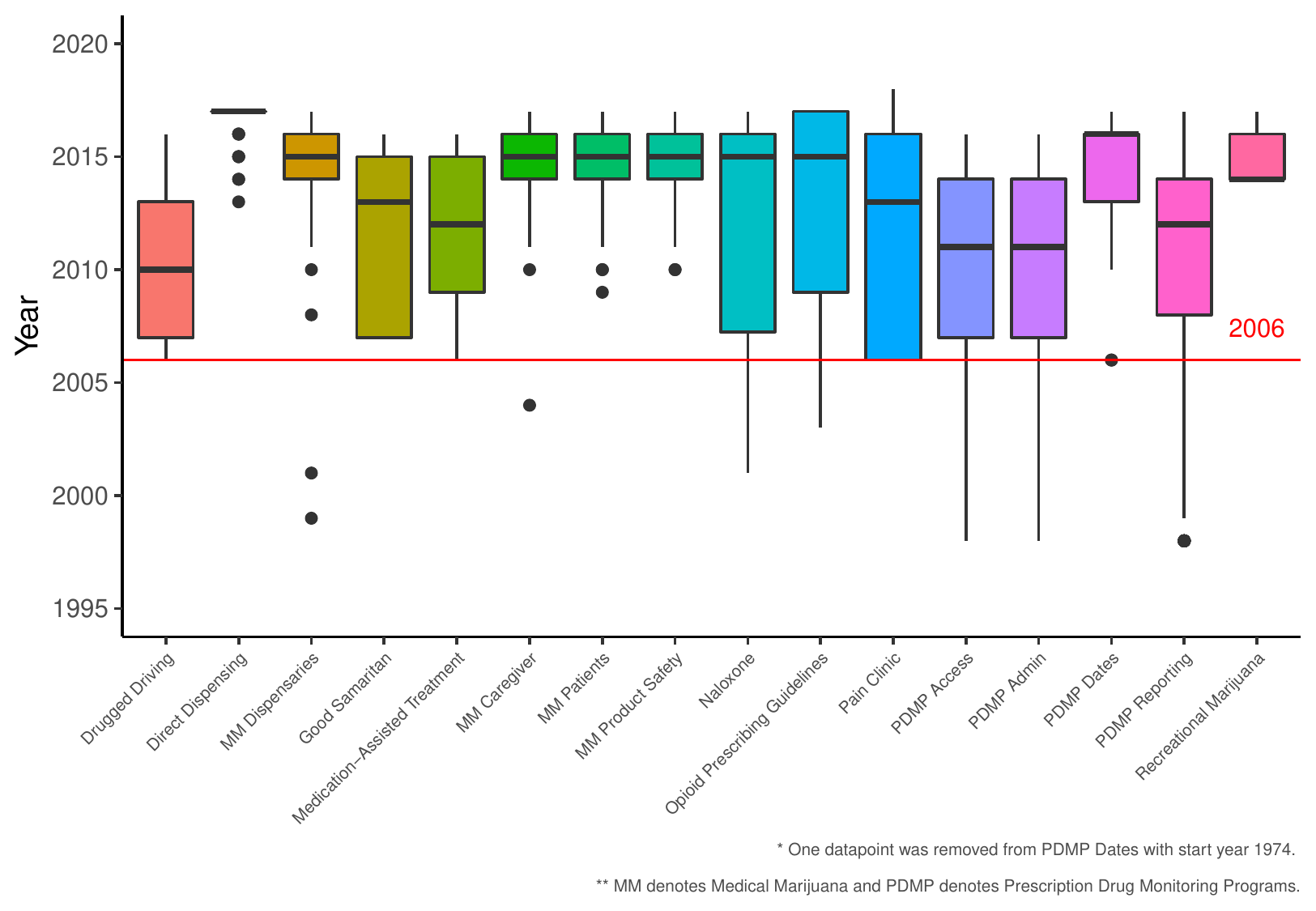}
\caption{\label{fig:PDAPS}Policy Start Years by PDAPS Dataset.}
\end{figure}

PDAPS is a data set to track key state laws related to prescriptions drug abuse, which provides a source of rigorous legal data to compare policies across time and states (Legal Science, 2018). The data for this analysis is broken down into 16 topic areas (data sets) including topics like Expanded Access to Naloxone, Good Samaritan 911 Immunity, Medical Marijuana, Opioid Related Controls, Prescription Drug Monitoring Program, and others. Generally, the data in PDAPS are a set of time-varying binary variables (Yes/No questions) to capture if a law was implemented or not for each state at a specific period of time. There are also some variables with multiple values nested to those binary variables for providing state-specific details. All binary variables in PDAPS were extracted from 16 data sets resulting in 145 state-law variables. While PDAPS records nested sub-questions as missing data if the law in the main question was not implemented, here the missing data is converted to ``No'' to represent the truth of fact that those nested-laws were also not implemented. Figure \ref{fig:PDAPS} displays the distribution of starting time of laws to demonstrate that although the first implemented state law for drug abuse can be traced back to 1974, most laws (more than 90 percent) were implemented after 2006. Therefore, even though WONDER provides data for opioid overdose death from 1999 to 2016, this paper focuses on the effects of drug laws on opioid overdose death rate from 2006 to 2016.

\begin{figure}
\centering
\includegraphics{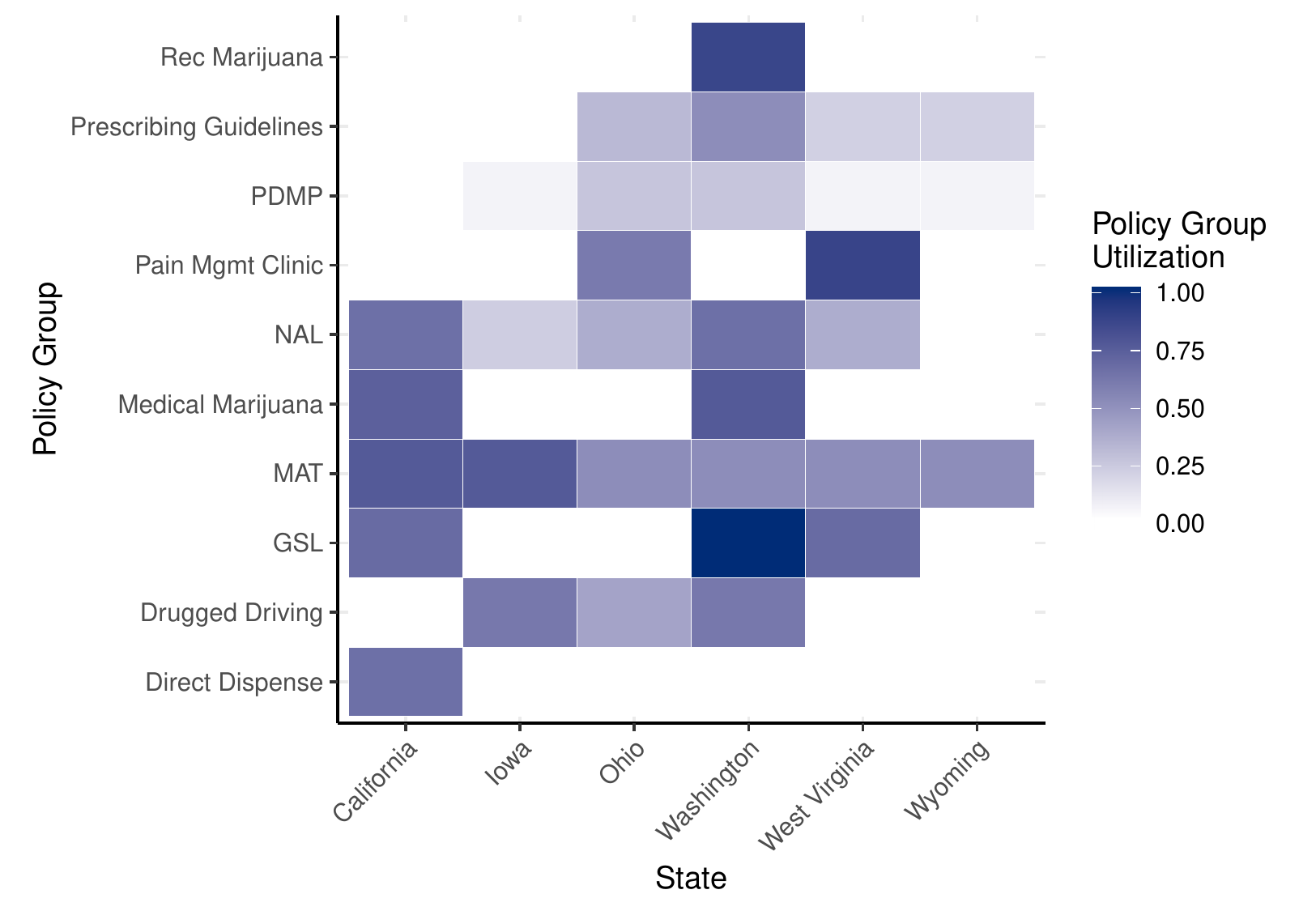}
\caption{\label{fig:showStateLaws}Selected States and Policy Utilization by Policy Group.}
\end{figure}

Figure \ref{fig:showStateLaws} shows a summary of the policy data for a selection of states. It summarizes the data into ``policy groups'' from PDAPS, which are sets of policies addressing a certain policy area (e.g.~NAL). These groups are different from the clusters to be used in this analysis. This shows the range of implementation, comparing a low-population state like Wyoming and a high-population state like California, or a state struggling with extremely high overdose death rates like Ohio. Darker colors indicate that the state has implemented closer to 100 percent of the policies within the group.

A common approach to utilize PDAPS data is to select a small-set of binary variables as explanatory variables for fitting a (Poisson) regression model to determine if the implemented laws are associated with changes in opioid-related overdose deaths, (e.g.~Bachhuber, Saloner, Cunningham, \& Barry, 2014; Patrick, Fry, Jones, \& Buntin, 2016; Rees, Sabia, Argys, Latshaw, \& Dave, 2017). However, as noted by Pardo (2017) and Fink et al. (2018), this approach may face problems of multicollinearity, inadequate adjustment for competing laws, and subjectivity in variable selection, all of which increase the risk of bias for studies. To avoid multicollinearity problems, Pardo (2017) created a score function to summarize prescription drug monitoring programs; however, this approach is still subjective in assigning different weights to administrative laws (Fink et al., 2018) and, perhaps more importantly, it loses the information of the effects on specific laws.

Several machine learning techniques were considered for reducing the dimension of PDAPS data to a readable level without losing much information, but most of them have shortcomings. Unsupervised learning methods (e.g., Principal Component Analysis (PCA) or Sparse PCA) were used to create clusters of laws, but the meaning of those clusters was difficult to interpret. For example, what does it mean for a linear combination of several prescriptions drug abuse related laws? A regression using a regularization technique, i.e., Lasso regression, was also tested for selecting a subset of important explanatory variables for opioid overdose death rate. However, neither fitting a Lasso regression using laws from all 16 data sets in PDAPS nor fitting several Lasso regressions only using the laws in each data set can efficiently reduce dimensionality. This may indicate that the sparsity assumption in Lasso regression (only a small number of laws may actually be relevant to the opioid overdose deaths) is violated.

This paper creates policy bundles using hierarchical clustering on state and year direction of the data. Instead of clustering on laws, this approach clusters the state-years with similar laws implemented. After a suitable number of feature clusters are extracted, the corresponding feature cluster for each state and year data can be identified. Then, those cluster features will be treated as explanatory variables to estimate the association between laws and opioid-related overdose deaths. The two main advantages of the proposed approach are as follows: First, feature clusters were created without sub-setting or combining any laws, increasing the ability to explain feature clusters. Second, this data-driven approach avoids the subjective bias seen frequently in the literature for selecting a particular subset of laws or assigning weights to laws.

\begin{figure}
\centering
\includegraphics{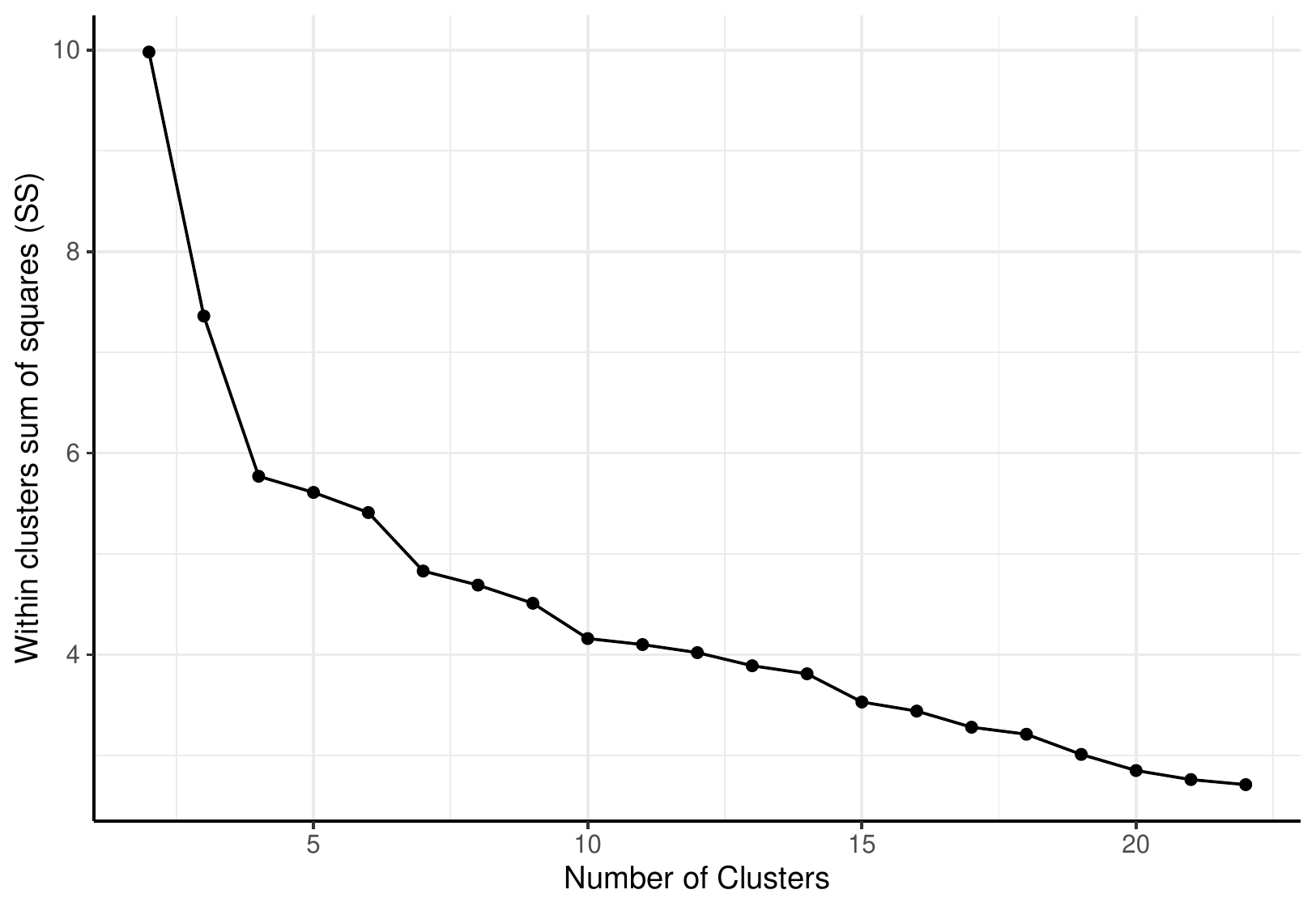}
\caption{\label{fig:elbowPlot}Elbow Plot for Hierarchical Clustering.}
\end{figure}

To perform hierarchical clustering on PDAPS data, several cleaning steps data were required. To facilitate eventual combination with overdose death rates (which are reported by year and state), the PDAPS data needed to be re-structured from the specific dates that the law was put into effect and repealed to ``years in force''. The effective date and valid through date for each state-law in PDAPS is used to identify the year for this study. If a particular state-law was implemented more than six months within a year, the law was treated as ``in effect'' for that year. Second, all laws in a binary variable form in PDAPS were selected except four policies with missing data and three laws without variation across the United States from 2006 to 2016, which results in \textbf{138 binary variables in total}. Table \ref{tab:hcDataInitialPrep} shows the policies removed from consideration. Then, the distance matrix was calculated using Gower distance (Gower, 1971), and the hierarchical clustering was performed with complete linkage (Maechler, Rousseeuw, Struyf, Hubert, \& Hornik, 2018). Finally, 10 feature clusters were selected based on the result of elbow method. The elbow method recommends thresholds for the number of clusters based on a decrease in the slope of the explainability of the clustering model for each threshold (Zambelli, 2016). Figure \ref{fig:elbowPlot} shows the elbow plot used for the decision. Because four clusters is the first major change in slope, and at 20, there is highly diminishing marginal change, those cases will be considered as sensitivity cases. All of these algorithms were executed in R (R Core Team, 2018).

\begin{table}

\caption{\label{tab:hcDataInitialPrep}Policies removed from clustering.}
\centering
\begin{tabular}[t]{ll>{\raggedright\arraybackslash}p{3in}}
\toprule
Variable name & Cause & Description\\
\midrule
DD\_PA & Missingness & Are Physician Assistants authorized to directly dispense controlled substances\\
DD\_PAlimit & Missingness & Are Physician Assistants limited to dispensing medications to the extent the supervising physician authorizes\\
DD\_PArur & Missingness & Are practitioners regulated differently based on the geographic location of their office\\
share.dispecheck & Missingness & Does the state require dispensers to check the PDMP before dispensing controlled substances\\
OPG\_.Acute.Penalty & Fixed & Is there a mandatory penalty for failure to comply with the acute pain guidelines\\
\addlinespace
OPG\_EDPenalty & Fixed & Is there a mandatory penalty for failure to comply with the emergency department guidelines\\
naloxone.crimpossessionprog & Fixed & Is participation in a naloxone administration program required as a condition of immunity\\
\bottomrule
\end{tabular}
\end{table}

\hypertarget{other-variables}{%
\subsection{Other Variables}\label{other-variables}}

Several time-varying demographic measures (confounding factors) are also considered, including Gini index, median household income, and opioid prescribing rate by state and year, and proportion of adults aged more than 25 years by state, year, and gender, where GINI index and household income were obtained from American Community Survey (ACS) in United State Census Bureau, prescribing rate was measured by Legal Science, LLC, and proportion of adults was derived from 1-year ACS data and estimated using table of age and sex in ACS five-year Estimates for each year, state, and gender. These variables are often suggested as relevant variables for opioid overdose death rate in literature (see for example Bachhuber et al., 2014; Pardo, 2017; Rees et al., 2017).

\hypertarget{results}{%
\section{Results}\label{results}}

\begin{figure}
\centering
\includegraphics{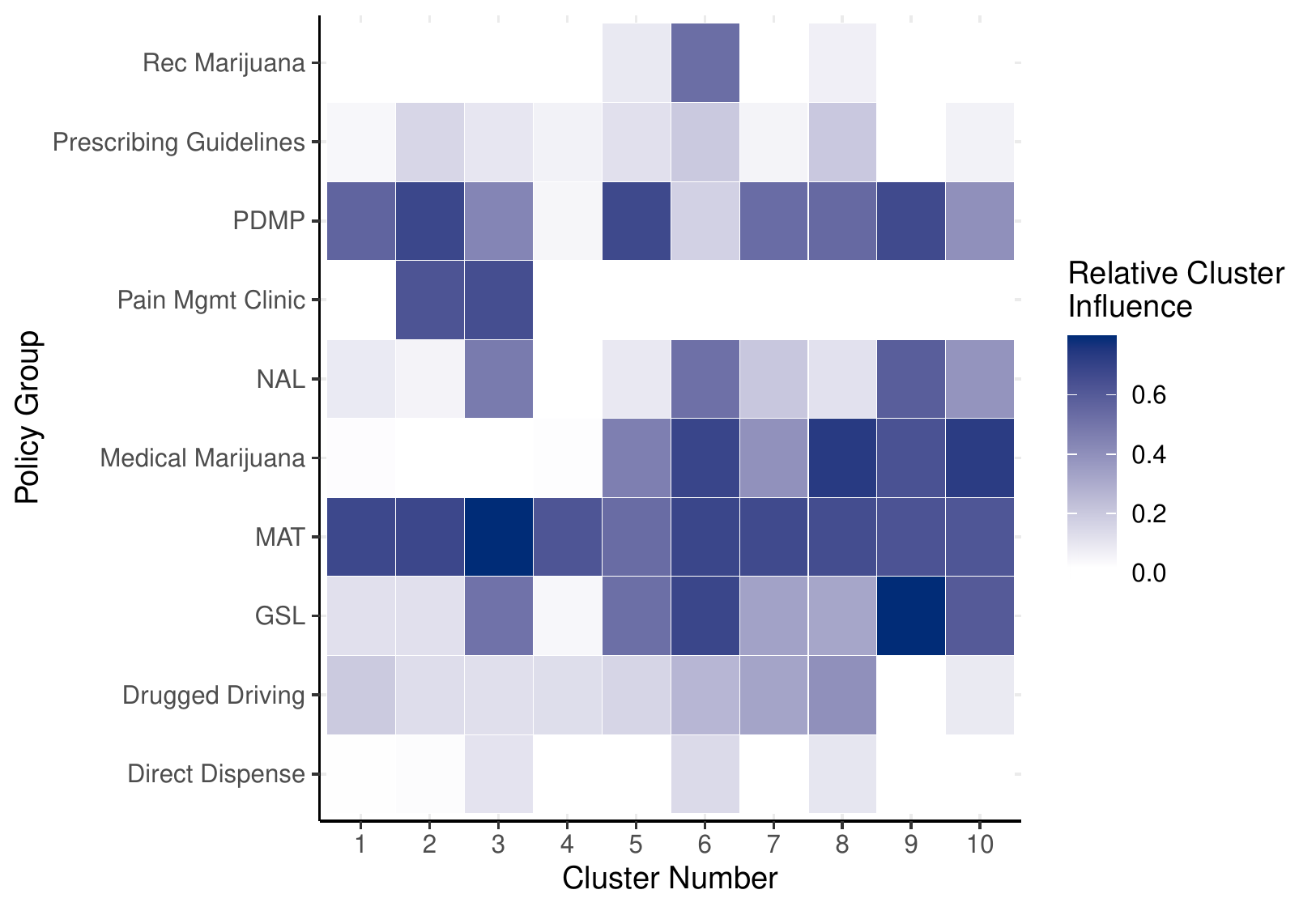}
\caption{\label{fig:HCPlot2}Policy Group influence on Cluster Definition.}
\end{figure}

The clustering into 10 clusters generated a data set that was used in the regression. As mentioned, each cluster contains several state-year data across 138 law variables. These policies are grouped in PDAPS by Legal Science (2018) and we call those groups ``Policy Groups'' for this analysis. As a summary of the clusters, Figure \ref{fig:HCPlot2} shows the relative effect of each policy group on the cluster definitions. Because of the limited space, the figure only shows the corresponding name of topic area for each law. The list of full policy questions in the figure can be found in the online appendix\footnote{All appendices are available by request from the corresponding author}. Using the figure, the key features of the 10 clusters and the number of observation (state and years) in each cluster are summarized in Table \ref{tab:HCTable}. More detailed plots, showing each policy group's full set of policies are available in the online appendix\footnote{All appendices are available by request from the corresponding author}.

\begin{table}

\caption{\label{tab:HCTable}Cluster descriptions and counts.}
\centering
\begin{tabular}[t]{rr>{\raggedright\arraybackslash}p{5in}}
\toprule
Cluster & N & Description\\
\midrule
1 & 298 & Represents early stage where a small set of laws was implemented, not including Medication-Assisted Treatment Laws and PDMP Laws.\\
2 & 39 & Similar to Cluster 1, but more PDMP laws were implemented and Pain Management Clinical laws were implemented.\\
3 & 17 & Similar to cluster 2, but more focus on Pain Management Clinical laws and MAT.\\
4 & 105 & Cluster with fewest laws, mostly just MAT.\\
5 & 33 & Focused on PDMP, with medical marijuana, MAD, and GSL focus.\\
\addlinespace
6 & 9 & Similar to cluster 5, but with less focus on PDMP, and adding recreational marijuana laws.\\
7 & 20 & Similar to cluster 6, bit no recreational marijuana laws and generally fewer laws.\\
8 & 11 & Similar to cluster 7, but adding more laws from Medical Marijuana, prescribing guidelines, and direct dispensing.\\
9 & 9 & Heavily focused on GSL, PDMP, NAL, Medical Marijuana and MAT.\\
10 & 20 & Similar to cluster 9, but lower focus on GSL, PDMP, and NAL, with a heavier focus on Medical Marijuana.\\
\bottomrule
\end{tabular}
\end{table}

The results shown in table \ref{tab:showBaseModels} allow us to reject the null hypothesis that the policy cluster do not affect drug overdose deaths. The fixed effects model (\(Year\) and \(State\)) shows that there is a statistical effect from including the policy clusters. Specifically, the log-likelihood of the model improves and the Aikake Information Criterion (AIC) decreases indicating better predictiveness without sacrificing efficiency. Because of the fixed effects, the table only shows the relative difference between clusters two to 10 and cluster one. The results omit the State and Year coefficients for simplicity. The effect of Gender does not change from model to model, but the effect of \(Prescribing.Rate\) is shown to be inefficient and biased (changing coefficient direction and becoming insignificant). \(Gini\) is moderated and \(Income\) is slightly moderated (recalling that \(Income\) has been scaled to the thousands). Clusters five, seven, and 10 all show an increased number of overdose deaths when implemented, compared to cluster 1.

The impact of a policy is possibly diffused or distributed over multiple years. To consider the sensitivity of the model, different time lags are included to show a policy's impact over time. In order to save space, Table \ref{tab:showBaseModels} omits the additional lag values after the causal lag (1 year), depending on the model used. In the model with just a two year lag, the AIC and log likelihood worsen, suggesting that the immediate effect is important to capture. Model three shows a five-year dissipating impact, and has the largest reduction in AIC.

\begin{table}[!htbp] \centering 
  \caption{Regression Model Results for overdose deaths.} 
  \label{tab:showBaseModels} 
\begin{tabular}{@{\extracolsep{5pt}}lcccc} 
\\[-1.8ex]\hline 
\hline \\[-1.8ex] 
 & \multicolumn{4}{c}{\textit{Dependent variable:}} \\ 
\cline{2-5} 
\\[-1.8ex] & \multicolumn{4}{c}{Deaths} \\ 
 & Base & 10Clust & 10Clust-2Yr & 10Clust-Dissipate \\ 
\\[-1.8ex] & (1) & (2) & (3) & (4)\\ 
\hline \\[-1.8ex] 
 GenderMale & 0.669$^{***}$ & 0.670$^{***}$ & 0.672$^{***}$ & 0.664$^{***}$ \\ 
  & (0.004) & (0.004) & (0.004) & (0.004) \\ 
  Prescribing.Rate & 0.004$^{***}$ & $-$0.0002 & 0.001$^{**}$ & $-$0.001$^{***}$ \\ 
  & (0.0004) & (0.0005) & (0.0005) & (0.0005) \\ 
  Gini & $-$2.684$^{***}$ & $-$2.131$^{***}$ & $-$0.231 & $-$1.283$^{**}$ \\ 
  & (0.516) & (0.538) & (0.541) & (0.551) \\ 
  Income & $-$0.003$^{***}$ & $-$0.002$^{***}$ & $-$0.002$^{**}$ & $-$0.004$^{***}$ \\ 
  & (0.001) & (0.001) & (0.001) & (0.001) \\ 
  Deaths.lag1 & 0.0001$^{***}$ & 0.0001$^{***}$ & 0.0001$^{***}$ & 0.0001$^{***}$ \\ 
  & (0.00001) & (0.00001) & (0.00001) & (0.00001) \\ 
  factor(cluster.lag1)2 &  & $-$0.125$^{***}$ &  & $-$0.072$^{***}$ \\ 
  &  & (0.009) &  & (0.013) \\ 
  factor(cluster.lag1)3 &  & $-$0.102$^{***}$ &  & $-$0.055$^{***}$ \\ 
  &  & (0.010) &  & (0.015) \\ 
  factor(cluster.lag1)4 &  & $-$0.042$^{***}$ &  & $-$0.010 \\ 
  &  & (0.008) &  & (0.011) \\ 
  factor(cluster.lag1)5 &  & 0.029$^{***}$ &  & 0.010 \\ 
  &  & (0.010) &  & (0.013) \\ 
  factor(cluster.lag1)6 &  & $-$0.136$^{***}$ &  & $-$0.038$^{**}$ \\ 
  &  & (0.013) &  & (0.015) \\ 
  factor(cluster.lag1)7 &  & 0.077$^{***}$ &  & 0.129$^{***}$ \\ 
  &  & (0.012) &  & (0.016) \\ 
  factor(cluster.lag1)8 &  & $-$0.073$^{***}$ &  & $-$0.040$^{*}$ \\ 
  &  & (0.016) &  & (0.024) \\ 
  factor(cluster.lag1)9 &  & $-$0.137$^{***}$ &  & 0.015 \\ 
  &  & (0.011) &  & (0.017) \\ 
  factor(cluster.lag1)10 &  & 0.221$^{***}$ &  & 0.286$^{***}$ \\ 
  &  & (0.010) &  & (0.013) \\ 
 \hline \\[-1.8ex] 
Observations & 1,121 & 1,121 & 1,120 & 1,117 \\ 
Log Likelihood & $-$10,448.770 & $-$9,829.777 & $-$9,957.005 & $-$9,324.868 \\ 
Akaike Inf. Crit. & 21,029.540 & 19,809.550 & 20,064.010 & 18,871.740 \\ 
\hline 
\hline \\[-1.8ex] 
\textit{Note:}  & \multicolumn{4}{r}{$^{*}$p$<$0.1; $^{**}$p$<$0.05; $^{***}$p$<$0.01} \\ 
\end{tabular} 
\end{table}

\begin{figure}
\centering
\includegraphics{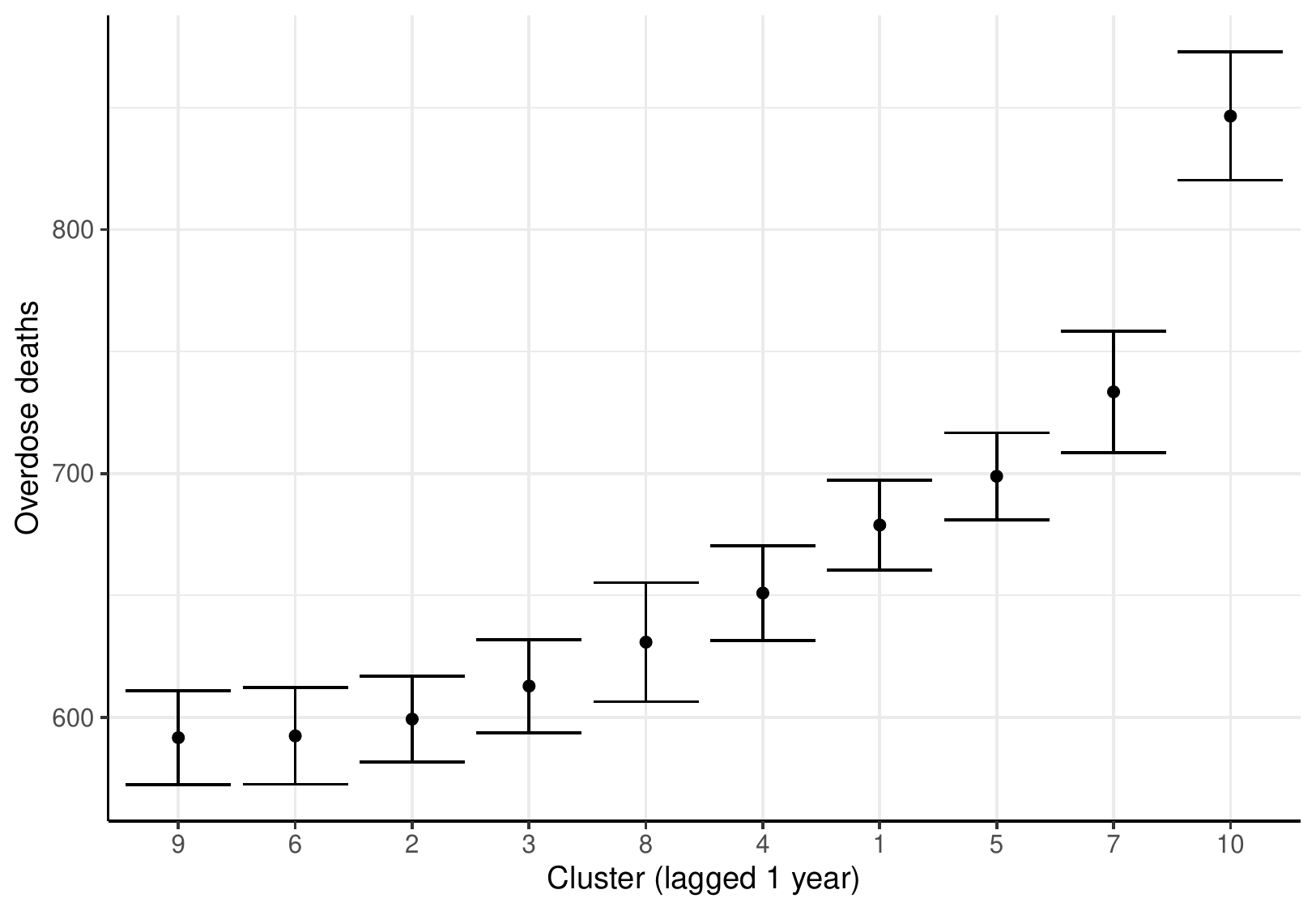}
\caption{\label{fig:showEffects}Relative Effects of Policy Clusters with \(2*SE\) Error Bars.}
\end{figure}

Figure \ref{fig:showEffects} shows the relative impacts of the different results visually, allowing for comparison of the policies as a collection. It is estimated by generating a synthetic data set holding all factors equal and varying only the cluster. The figure shows that differences between clusters are sometimes minimally consequential. However, the difference between the best performing cluster (cluster nine) and the worst performing (cluster 10) is significant (confidence intervals do not overlap) and substantial (approximately a \(30\%\) decrease).

\begin{figure}
\centering
\includegraphics{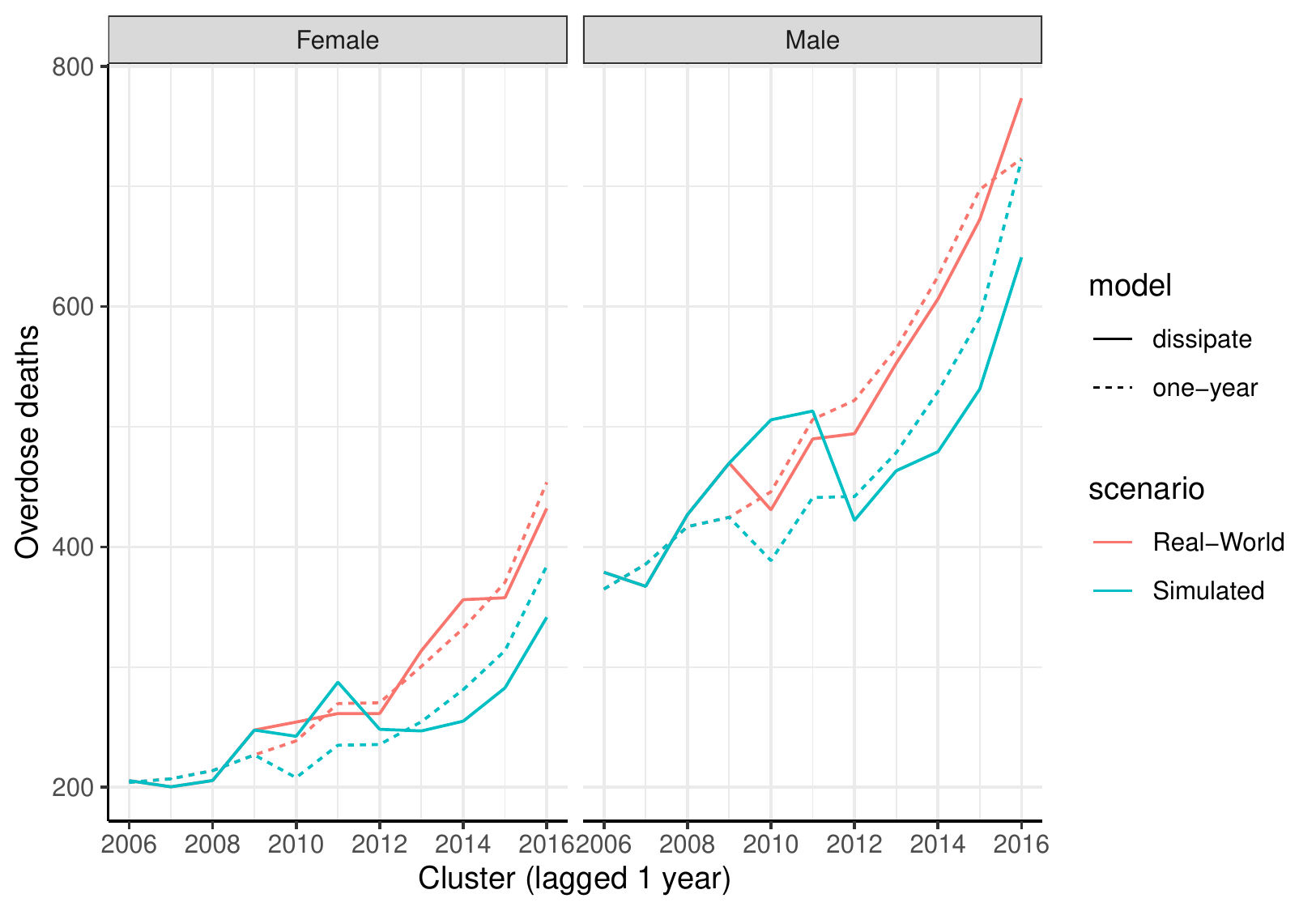}
\caption{\label{fig:dissipatingSim}Simulated Impact of Change to Policy 9 in Washington, 2010.}
\end{figure}

The impact of the difference in effectiveness in policy for a specific state, like Washington, is substantial. Assuming a change in the year 2010, Figure \ref{fig:dissipatingSim} shows the potential change in trajectory for both the single lag and the dissipating models for a world where cluster nine policies are implemented (``simulated'') vs.~not (``real-world''). The simulated results show an immediate interruption for the one-year lag model and a one-year delayed impact in the five-year dissipating model (illustrating the ``interrupted time-series design''). The figure also shows the complex interactions between environmental factors and policy change outcomes for the different genders. The differences for men are larger than for women. The figure compares the predicted change in Washington of the policies with the multiyear shock model, and the effect is slightly larger, but also delayed compared to a single lag. with the multiyear model, but model performance increases substantially, suggesting that this is a more reliable prediction.

To rule out selection bias from our selection of the number of clusters, Table \ref{tab:diffClusterN} compares the effects of different numbers of clusters. The 20 cluster model has the highest model performance, but as discussed above, is somewhat more complicated to explain.

\begin{table}[!htbp] \centering 
  \caption{Sensitivity regression model results.} 
  \label{tab:diffClusterN} 
\begin{tabular}{@{\extracolsep{5pt}}lcccc} 
\\[-1.8ex]\hline 
\hline \\[-1.8ex] 
 & \multicolumn{4}{c}{\textit{Dependent variable:}} \\ 
\cline{2-5} 
\\[-1.8ex] & \multicolumn{4}{c}{Deaths} \\ 
 & 10Clust & 4Clust & 20Clust & 20Clust-dissipate \\ 
\\[-1.8ex] & (1) & (2) & (3) & (4)\\ 
\hline \\[-1.8ex] 
 GenderMale & 0.670$^{***}$ & 0.673$^{***}$ & 0.669$^{***}$ & 0.661$^{***}$ \\ 
  & (0.004) & (0.004) & (0.004) & (0.005) \\ 
  Prescribing.Rate & $-$0.0002 & 0.001$^{**}$ & $-$0.0003 & $-$0.001 \\ 
  & (0.0005) & (0.0005) & (0.0005) & (0.001) \\ 
  Gini & $-$2.131$^{***}$ & $-$4.021$^{***}$ & $-$2.470$^{***}$ & $-$1.187$^{**}$ \\ 
  & (0.538) & (0.527) & (0.547) & (0.578) \\ 
  Income & $-$0.002$^{***}$ & $-$0.005$^{***}$ & $-$0.002$^{***}$ & $-$0.004$^{***}$ \\ 
  & (0.001) & (0.001) & (0.001) & (0.001) \\ 
  Deaths.lag1 & 0.0001$^{***}$ & 0.0001$^{***}$ & 0.0001$^{***}$ & 0.0001$^{***}$ \\ 
  & (0.00001) & (0.00001) & (0.00001) & (0.00001) \\ 
 \hline \\[-1.8ex] 
Observations & 1,121 & 1,121 & 1,121 & 1,117 \\ 
Log Likelihood & $-$9,829.777 & $-$10,333.000 & $-$9,677.593 & $-$8,756.953 \\ 
Akaike Inf. Crit. & 19,809.550 & 20,803.990 & 19,525.190 & 17,835.910 \\ 
\hline 
\hline \\[-1.8ex] 
\textit{Note:}  & \multicolumn{4}{r}{$^{*}$p$<$0.1; $^{**}$p$<$0.05; $^{***}$p$<$0.01} \\ 
\end{tabular} 
\end{table}

\hypertarget{discussion}{%
\section{Discussion}\label{discussion}}

This study shows policy is an important variable in the comparative state analysis of drug overdose deaths and hierarchical clustering can illustrate the relative impact of policies on the change in drug overdose deaths. Hierarchical clusters use empirical, and reproducible methods to segment individual policies that are highly correlated into bundles, and those policy bundles can have a highly differentiated effect. The sensitivity analysis suggests there are major challenges in selecting the ``correct'' number of clusters, and also that policy has a changing effect over time.

\begin{figure}
\centering
\includegraphics{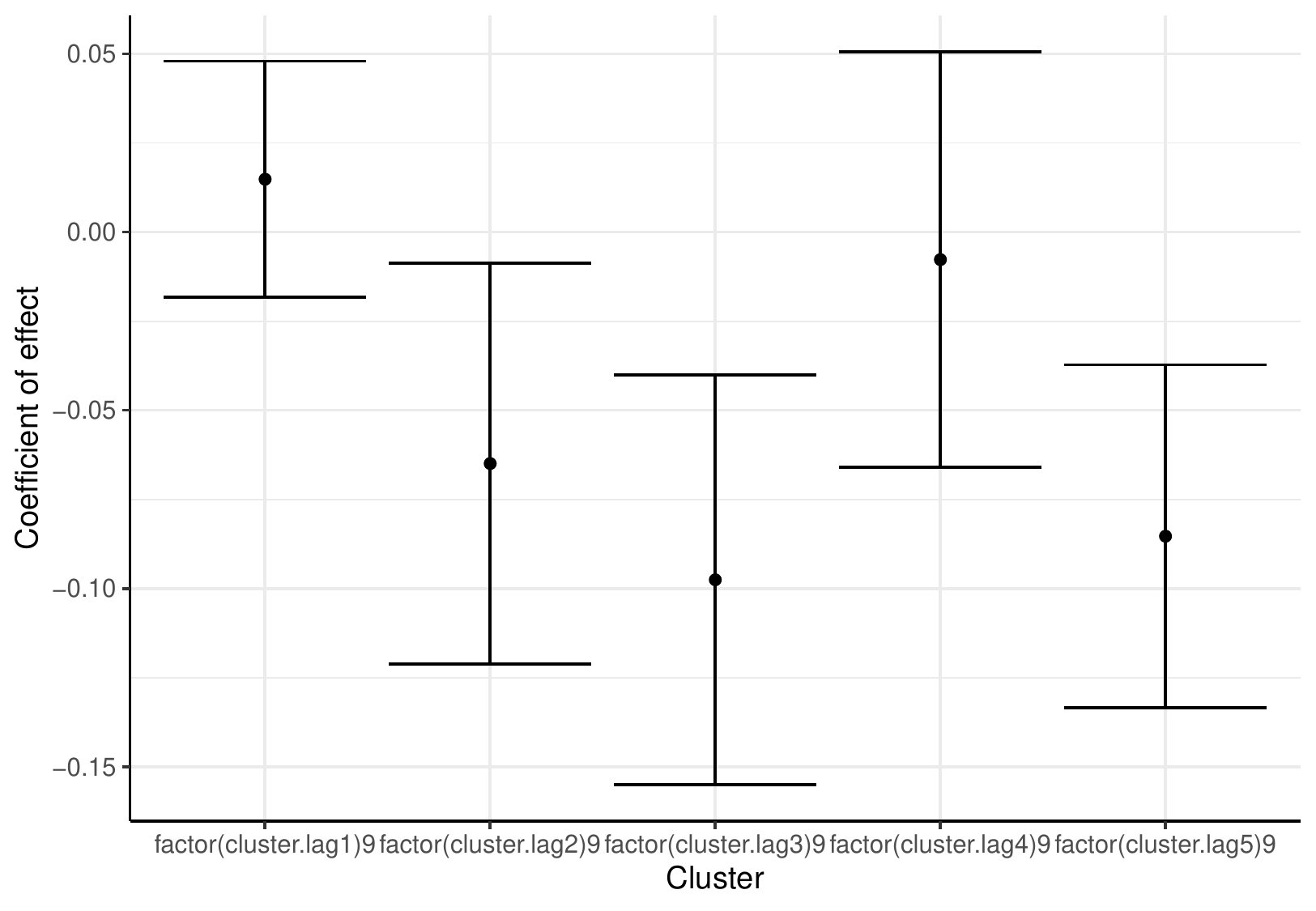}
\caption{\label{fig:showAttenuation}Attenuation of Policy Impact Over Time for Cluster nine.}
\end{figure}

This changing effect over time fits an assumption where behavior adapts to new policies and the policy impacts are eventually muted. This learning process necessitates continual innovation in policy to show continued growth. As an illustration of this learning, Figure \ref{fig:showAttenuation} shows the relative effect of policy bundle nine from the 10 policy bundle over five years. In this case, the initial neutral effect in the first year ultimately leads to a reduction in drug overdoses in years two and 3, with a neutral impact thereafter. Controlling for time reduces the likelihood that this pattern is simply the long term trend toward increased opioid overdose deaths.

Based on these results, policy bundle nine has the largest effect. There are 18 state-years matching that bundle (California and New Mexico, ranging from 2011-2015). Of those state-years, there are 74 total policies in force, with 30 implemented in all state-years. Table \ref{tab:digInto9AllMatch} shows the policies that were in place in all states for cluster nine (30 policies). This cluster is dominated by GSL, PDMP, NAL, MAT, and Medical Marijuana policies. There is balance in the legalization, treatment, and protection from prosecution, with heavy emphasis on access to medical marijuana.

\begin{longtable}[t]{>{\raggedright\arraybackslash}p{6in}}
\caption{\label{tab:digInto9AllMatch}Policies in cluster nine in force in all state-years.}\\
\hline
Description\\
\hline
Does the law allow in-state law enforcement to access PDMP data\\
\hline
Is treatment planning for admitted patients required at OTPs\\
\hline
Is periodic review of the treatment plan required by law\\
\hline
Does the law explicitly provide any affirmative defenses for the use of medical marijuana\\
\hline
Does the state allow qualifying patients to designate a caregiver for assistance in the use of medical marijuana\\
\hline
Are caregivers allowed to cultivate marijuana\\
\hline
Does the state have a law authorizing adults to use medical marijuana\\
\hline
Are registered caregivers who comply with the law exempt from arrest\\
\hline
Are caregivers allowed to transport medical marijuana\\
\hline
Does the law explicitly allow cardholders to cultivate medical marijuana plants\\
\hline
Does the state have a law regulating medical marijuana dispensaries\\
\hline
Are dispensaries required to operate as not-for-profit entities\\
\hline
Does the state law provide an explicit procedure for adding additional qualifying diseases\\
\hline
Are minors authorized to use medical marijuana\\
\hline
Does the state law have explicit privacy provisions related to medical marijuana cardholders\\
\hline
Does the jurisdiction have a naloxone access law\\
\hline
Do prescribers have immunity from civil liability for prescribing dispensing or distributing naloxone to a layperson\\
\hline
Do prescribers have immunity from criminal prosecution for prescribing, dispening or distributing naloxone to a layperson\\
\hline
Is a layperson immune from criminal liability when administering naloxone\\
\hline
Do dispensers have immunity from civil liability for prescribing, dispensing or distributing naloxone to a layperson\\
\hline
Do dispensers have immunity from criminal prosecution for prescribing, dispensing or distributing naloxone to a layperson\\
\hline
Are prescribers required to act with reasonable care\\
\hline
Does the law explicitly allow out-of-state law enforcement to access PDMP data\\
\hline
Does this state have legislation authorizing access by professionals to a PDMP system\\
\hline
Does this state have legislation authorizing a PDMP\\
\hline
Does the state have legislation authorizing a PDMP\\
\hline
Is the PDMP permitted or required to identify suspicious or statistically outlying prescribing dispensing or purchasing activity\\
\hline
Does the law permit or require PDMP to release de-identified data for research or education\\
\hline
Does this state have legislation requiring dispensers to report data to the PDMP\\
\hline
\end{longtable}

\begin{figure}
\centering
\includegraphics{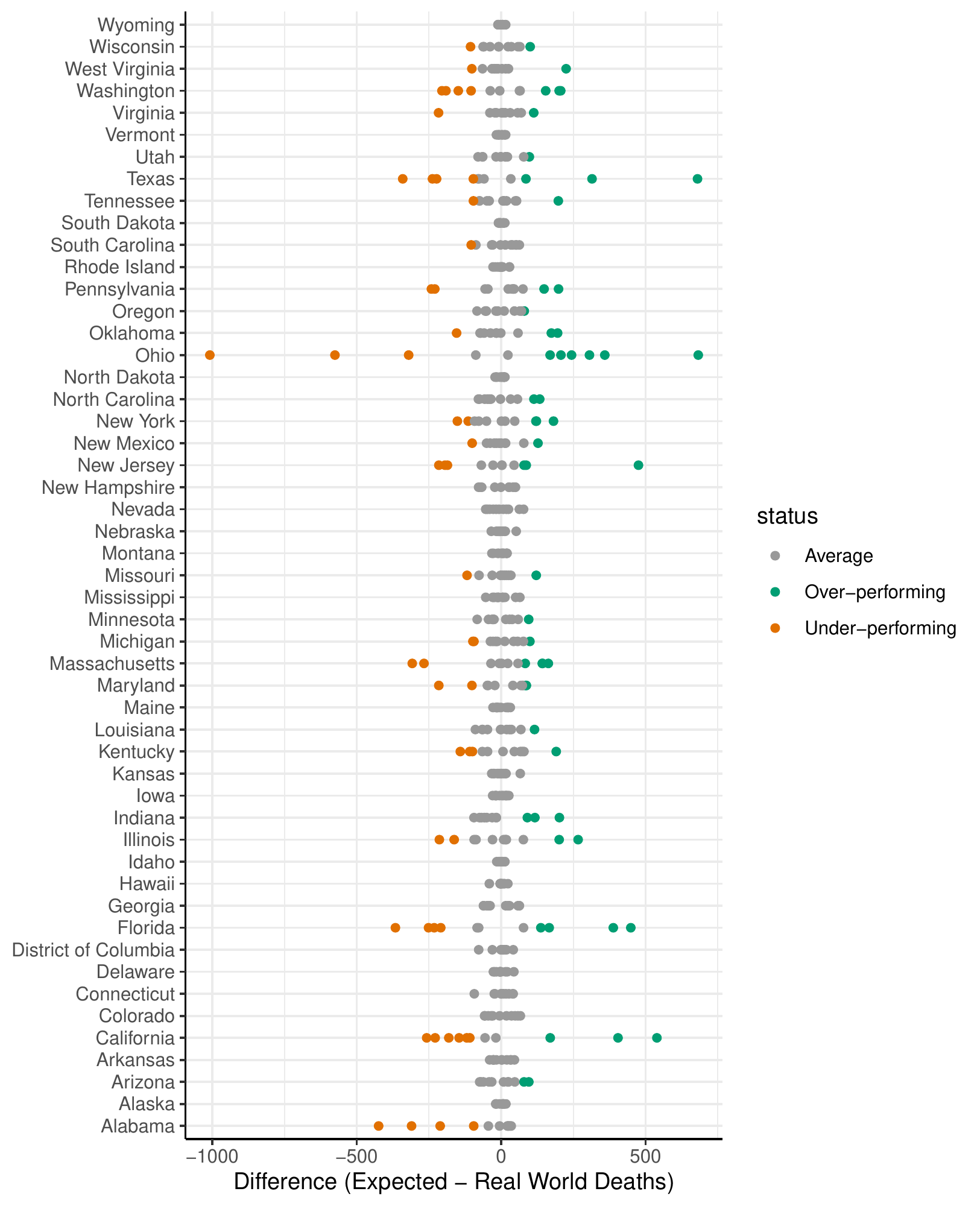}
\caption{\label{fig:brightspots}Bright Spot Analysis for Drug Overdose Deaths.}
\end{figure}

These analytical tools can also help understand states that are outperforming (often called ``bright spots'') and states that are under performing their expectations. These states (in a specific time-frame) can be useful in case studies to help drive new research. Detailed qualitative CPA can identify potential hypotheses on a policy bundle's overall impact that can then feed a virtuous cycle of research. Given the robust documentation of the policy environments in each state, this could lead to new policy innovations and evaluations. This also provides an opportunity to identify details about policies as implemented, compared to policies as stated. The critical importance of this difference was highlighted by Ritter et al. (2016). Figure \ref{fig:brightspots} shows a simple analysis showing some of these states.

\hypertarget{limitations}{%
\subsection{Limitations}\label{limitations}}

This study and approach produced new insight into the effectiveness of the policy system, illustrating how policies can be complementary and supportive or contradictory and destructive at reducing drug overdose deaths. There are limitations that should be addressed for future research into both the conclusions of the study and the clustering approach used to generate policy bundles.

Future research should address the specific cause of death more directly as an outcome and the effect of policy bundles on those death rates to provide more detailed information to policymakers. While opioid overdose is the leading cause of drug overdose deaths, other drug overdose deaths are incorporated in this publicly-available dataset. The policies in this case are not limited to just opioid overdoses, but primarily focus on those policies. Future research using more detailed data, some of which might not be publicly available would improve those results.

This study also fails to directly model an important predictor of drug overdoses as an independent variable: drug use (a behavior change). NSDUH data would be a useful source for the relationship between policy and drug use, but handling the aggregation of that data for state-level analysis is a challenge beyond the scope of this paper.

While hierarchical clustering substantially reduced the complexity of the data set (from 138 policies to four, 10, or 20, depending on the scenario), the explainability of the bundles is critical to the finding's ability to be used in a policymaking context. Future work should leverage legal analysts and data visualization tools to best understand these trade offs. While the model with 20 bundles was the most predictive, it also suffered from potential over-fitting and explainability challenges. Selection of the ``right'' number of clusters should be considered in future research.

Finally, the quality of data available using publicly available national datasets like CDC WONDER is difficult to assess. The data is drawn from medical examiner reports on cause of death, and depend extensively on the quality of that observation and reporting. Additional research into the reliability of outcome reporting for drug overdose deaths would strengthen the conclusions that can be drawn when using it as an outcome variable.

\hypertarget{conclusion}{%
\section{Conclusion}\label{conclusion}}

While drug overdose deaths have skyrocketed to epidemic levels in the U.S. (Centers for Disease Control and Prevention, 2018) and policy innovation has attempted to keep pace, the ability of policymakers to identify the core sets of policies that will treat the epidemic has been limited by the methodologies of CPA. While CPA as a practice is quite diverse and useful for some policy analysis, qualitative case studies that are difficult to generalize and quantitative studies that either lack specificity (by not considering the complexity of policy), or lack generalizability (by not including the \emph{system} of policies) have only partially served those decisionmaking needs.

This study provides information for policymakers about the effectiveness of existing drug policies using a new method for treating policies as ``bundles'' empirically and estimating the relative effect of those policies. In a policymaking environment, we would propose that a suite of policies based on the most impactful bundles (for example, bundle nine with a heavy GSL, PDMP, NAL, MAT, and Medical Marijuana focus) would have the largest impact on addressing the current drug overdose epidemic. Evaluators and researchers can leverage this approach to control for collinearity in the policy system and provide more detailed results about the sets of policies that have the most impact.

\hypertarget{acknowledgements}{%
\section{Acknowledgements}\label{acknowledgements}}

This research was funded by NIH-SBIR Research through grant 2R44DA040340-02. The authors would like to express appreciation for extensive research assistance by Jordan Vasko and Steven Watanabe and the support of Dr.~Jeremy Bellay and Dr.~Heidi Grunwald initiating the project.

\hypertarget{references}{%
\section*{References}\label{references}}
\addcontentsline{toc}{section}{References}

\hypertarget{refs}{}
\leavevmode\hypertarget{ref-bachhuber_medical_2014}{}%
Bachhuber, M. A., Saloner, B., Cunningham, C. O., \& Barry, C. L. (2014). Medical cannabis laws and opioid analgesic overdose mortality in the United States, 1999-2010. \emph{JAMA Internal Medicine}, \emph{174}(10), 1668--1673. \url{https://doi.org/10.1001/jamainternmed.2014.4005}

\leavevmode\hypertarget{ref-baptist_carrie_qualitative_2015}{}%
Baptist, Carrie, \& Befani, Barbara. (2015). Qualitative Comparative Analysis -- A Rigorous Qualitative Method for Assessing Impact Better Evaluation. Retrieved from \url{https://www.betterevaluation.org/en/resources/guide/qcr_a_rigorous_qualitative_method_for_assessing_impact}

\leavevmode\hypertarget{ref-bernal_interrupted_2017}{}%
Bernal, J. L., Cummins, S., \& Gasparrini, A. (2017). Interrupted time series regression for the evaluation of public health interventions: A tutorial. \emph{International Journal of Epidemiology}, \emph{46}(1), 348--355. \url{https://doi.org/10.1093/ije/dyw098}

\leavevmode\hypertarget{ref-blackman_using_2013}{}%
Blackman, T., Wistow, J., \& Byrne, D. (2013). Using Qualitative Comparative Analysis to understand complex policy problems. \emph{Evaluation}, \emph{19}(2), 126--140. \url{https://doi.org/10.1177/1356389013484203}

\leavevmode\hypertarget{ref-u.s._census_bureau_annual_2017}{}%
Bureau, U. S. C. (2017a). \emph{Annual Household Income: Households, 1 year Estimates, 2006-2016}.

\leavevmode\hypertarget{ref-u.s._census_bureau_gini_2017}{}%
Bureau, U. S. C. (2017b). \emph{GINI INDEX OF INCOME INEQUALITY: Households, 1 year Estimates, 2006-2016}. Retrieved from \url{https://factfinder.census.gov/bkmk/table/1.0/en/ACS/17_1YR/B19083/0100000US.04000}

\leavevmode\hypertarget{ref-burris_theory_2017}{}%
Burris, S. (2017). Theory and methods in comparative drug and alcohol policy research: Response to a review of the literature. \emph{International Journal of Drug Policy}, \emph{41}(Supplement C), 126--131. \url{https://doi.org/10.1016/j.drugpo.2016.11.011}

\leavevmode\hypertarget{ref-burris_making_2010}{}%
Burris, S., Wagenaar, A. C., Swanson, J., Ibrahim, J. K., Wood, J., \& Mello, M. M. (2010). Making the Case for Laws That Improve Health: A Framework for Public Health Law Research. \emph{Milbank Quarterly}, \emph{88}(2), 169--210. \url{https://doi.org/10.1111/j.1468-0009.2010.00595.x}

\leavevmode\hypertarget{ref-centers_for_disease_control_and_prevention_underlying_2017}{}%
Centers for Disease Control and Prevention, N. C. for H. S. (2017). \emph{Underlying Cause of Death 1999-2016 on CDC WONDER Online Database}. Retrieved from \url{http://wonder.cdc.gov/ucd-icd10.html}

\leavevmode\hypertarget{ref-centers_for_disease_control_and_prevention_understanding_2018}{}%
Centers for Disease Control and Prevention, N. C. for I. P. and C. (2018). \emph{Understanding the Epidemic Drug Overdose CDC Injury Center}. Retrieved from \url{https://www.cdc.gov/drugoverdose/epidemic/index.html}

\leavevmode\hypertarget{ref-devers_using_2016}{}%
Devers, K. J., Lallemand, N. C., Burton, R. A., Zuckerman, S., \& Authors, A. (2016). Using Qualitative Comparative Analysis (QCA) to Study Patient-Centered Medical Homes. Retrieved from \url{https://www.urban.org/research/publication/using-qualitative-comparative-analysis-qca-study-patient-centered-medical-homes}

\leavevmode\hypertarget{ref-centers_for_disease_control_and_prevention_us_2019}{}%
Disease Control and Prevention, C. for. (2019). \emph{US Opioid Prescribing Rate Maps Drug Overdose CDC Injury Center}. Retrieved from \url{https://www.cdc.gov/drugoverdose/maps/rxrate-maps.html}

\leavevmode\hypertarget{ref-fink_association_2018}{}%
Fink, D. S., Schleimer, J. P., Sarvet, A., Grover, K. K., Delcher, C., Castillo-Carniglia, A., \ldots{} Cerdá, M. (2018). Association Between Prescription Drug Monitoring Programs and Nonfatal and Fatal Drug Overdoses: A Systematic Review. \emph{Annals of Internal Medicine}, \emph{168}(11), 783--790. \url{https://doi.org/10.7326/M17-3074}

\leavevmode\hypertarget{ref-gower_general_1971}{}%
Gower, J. C. (1971). A general coefficient of similarity and some of its properties. \emph{Biometrics}, \emph{27}(4), 857--871. Retrieved from \url{http://links.jstor.org/sici?sici=0006-341X\%28197112\%2927\%3A4\%3C857\%3AAGCOSA\%3E2.0.CO\%3B2-3}

\leavevmode\hypertarget{ref-haegerich_what_2014}{}%
Haegerich, T. M., Paulozzi, L. J., Manns, B. J., \& Jones, C. M. (2014). What we know, and don't know, about the impact of state policy and systems-level interventions on prescription drug overdose. \emph{Drug and Alcohol Dependence}, \emph{145}(Supplement C), 34--47. \url{https://doi.org/10.1016/j.drugalcdep.2014.10.001}

\leavevmode\hypertarget{ref-keele_natural_2015}{}%
Keele, L., \& Titiunik, R. (2015). Natural Experiments Based on Geography. \emph{Political Science Research and Methods}, \emph{FirstView}, 1--31. \url{https://doi.org/10.1017/psrm.2015.4}

\leavevmode\hypertarget{ref-legal_science_llc_prescription_2018}{}%
Legal Science, L. (2018). \emph{Prescription Drug Abuse Policy System - PDAPS}. Retrieved from \url{http://pdaps.org/}

\leavevmode\hypertarget{ref-maechler_cluster_2018}{}%
Maechler, M., Rousseeuw, P., Struyf, A., Hubert, M., \& Hornik, K. (2018). \emph{Cluster - Cluster Analysis Basics and Extensions}.

\leavevmode\hypertarget{ref-pardo_more_2017}{}%
Pardo, B. (2017). Do more robust prescription drug monitoring programs reduce prescription opioid overdose? \emph{Addiction}, \emph{112}(10), 1773--1783. \url{https://doi.org/10.1111/add.13741}

\leavevmode\hypertarget{ref-patrick_implementation_2016}{}%
Patrick, S. W., Fry, C. E., Jones, T. F., \& Buntin, M. B. (2016). Implementation Of Prescription Drug Monitoring Programs Associated With Reductions In Opioid-Related Death Rates. \emph{Health Affairs (Project Hope)}, \emph{35}(7), 1324--1332. \url{https://doi.org/10.1377/hlthaff.2015.1496}

\leavevmode\hypertarget{ref-r_core_team_r_2018}{}%
R Core Team. (2018). \emph{R - A Language and Environment for Statistical Computing}. Retrieved from \url{https://www.R-project.org/}

\leavevmode\hypertarget{ref-rees_little_2017}{}%
Rees, D. I., Sabia, J. J., Argys, L. M., Latshaw, J., \& Dave, D. (2017). \emph{With a Little Help from My Friends: The Effects of Naloxone Access and Good Samaritan Laws on Opioid-Related Deaths} (Working Paper No. 23171). \url{https://doi.org/10.3386/w23171}

\leavevmode\hypertarget{ref-ritter_comparative_2016}{}%
Ritter, A., Livingston, M., Chalmers, J., Berends, L., \& Reuter, P. (2016). Comparative policy analysis for alcohol and drugs: Current state of the field. \emph{International Journal of Drug Policy}, \emph{31}(Supplement C), 39--50. \url{https://doi.org/10.1016/j.drugpo.2016.02.004}

\leavevmode\hypertarget{ref-zambelli_data-driven_2016}{}%
Zambelli, A. E. (2016). A data-driven approach to estimating the number of clusters in hierarchical clustering. \emph{F1000Research}, \emph{5}. \url{https://doi.org/10.12688/f1000research.10103.1}

\end{document}